\documentclass[prb,aps,twocolumn,amsmath,amssymb,floatfix,
superscriptaddress, raggedbottom]{revtex4-2}
\usepackage[dvips]{graphics}
\usepackage{amsmath}
\usepackage{physics}
\usepackage{color}
\usepackage{float}
\usepackage{natbib}
\usepackage[dvipsnames]{xcolor}
\setcitestyle{numbers}
\setcitestyle{square}
\definecolor{dred}{rgb}{0,0,0.6}
\usepackage{graphicx}
\usepackage{bm}  
\usepackage{amssymb}  
\usepackage{amsmath}
\usepackage{braket}
\usepackage[mathscr]{euscript}
\usepackage{mathtools}
\usepackage[colorlinks=true, allcolors=blue]{hyperref}
\begin{document}
\title{Quasiperiodic potential induced corner states in a quadrupolar insulator}

\author{Srijata Lahiri and Saurabh Basu \\ \textit{Department of Physics, Indian Institute of Technology Guwahati-Guwahati, 781039 Assam, India}}
\date{\today}
\begin{abstract} 
We systematically investigate the topological and localization properties of a quadrupolar insulator represented by the celebrated Benalcazar-Bernevig-Hughes model in presence of a quasiperiodic disorder instilled in its hopping amplitude.
While disorder can be detrimental to the existence of the topological order in a system, we observe the emergence of a disorder driven topological phase where the original (clean) system demonstrates trivial behavior.
This phenomenon is confirmed by the re-emergence of zero energy states in the bandstructure together with a non-zero bulk quadrupole moment, which in turn establishes the bulk boundary correspondence (BBC).
Furthermore, the distribution of the excess electronic charge shows a pattern that is reminiscent of the bulk quadrupole topology.
To delve into the localization properties of the mid-band states, we compute the inverse participation and normalized participation ratios.
It is observed that the in-gap states become critical (multifractal) at the point that discerns a transition from a topological localized to a trivial localized phase.
Finally, we carry out a similar investigation to ascertain the effect of the quasiperiodic disorder on the quadrupolar insulator when the model exhibits topological properties in the absence of disorder.
Again, we note a multifractal behavior of the eigenstates in the vicinity of the transition.
\end{abstract}
\maketitle
\section{Introduction}
Topological insulators (TI) represent an area of supreme interest in the field of condensed matter physics for almost over a decade, being the cradle of exotic phenomena like anomalous quantum Hall and quantum spin Hall phases \cite{hasan2010colloquium, qi2011topological, hasan2011three, moore2010birth}.
TI correspond to systems in $d$ dimension, that exhibit insulating properties in the bulk, while hosting robust metallic surface states at the $d-1$ dimensional boundary.
\begin{figure}
\includegraphics[width=0.8\columnwidth]{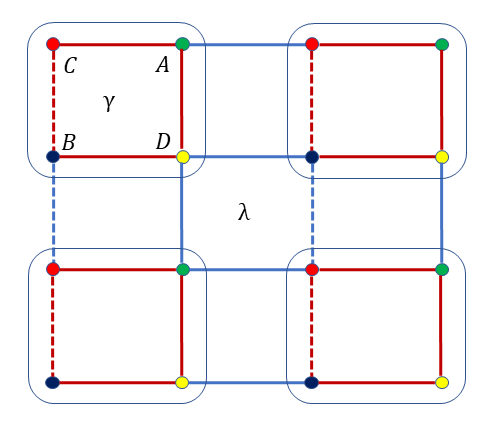}
\caption{\label{1}A schematic diagram of the Benalcazar-Bernevig-Hughes (BBH) model is shown. Each unit cell consists of four sublattices marked as $A$, $B$, $C$ and $D$. The intracellular hopping amplitude is shown by $\gamma$, whereas $\lambda$ represents the intercellular hopping. The dotted lines correspond to a negative hopping amplitude pertaining to a flux of magnitude $\pi$ that penetrates each plaquette.}
\end{figure}
The field of topology has been extensively explored both from a theoretical and experimental front, providing numerous material candidates that vividly exhibit robust boundary features. 
Examples include the seminal work by Bernevig \textit{et al.} and K\~onig \textit{et al.} on the quantum spin Hall effect in CdTe/HgTe quantum wells \cite{doi:10.1126/science.1133734, konig2007quantum}, followed by several works on Dirac and Weyl semimetals \cite{Deng2016, PhysRevLett.107.127205, PhysRevLett.107.186806, PhysRevB.96.125102}, topological superconductors \cite{PhysRevLett.110.117001, Ando_2013}, topological crystalline insulators \cite{Hsieh2012, Tanaka2012} etc.

Currently, higher order topological insulators (HOTI) have garnered considerable interest, as a noteworthy extension to the field of conventional first order TI \cite{schindler2018higher, khalaf2018higher, matsugatani2018connecting, franca2018anomalous}.
HOTI feature robust topological states at boundaries of dimension less than $d-1$ for a topologically non-trivial $d$-dimensional bulk, leading to the formation of corner states in two-dimensional (2D) and corner/hinge states in three-dimensional (3D) systems.
While experimental evidence of real material candidates featuring higher order states in 2D remain absent, evidence of second order topology has been found in 3D in Bismuth \cite{Schindler2018} and stacked Bismuth halide chains \cite{Noguchi2021}.

Topological insulators with quantized quadrupole moment require a special mention in the field of higher order topology.
The quadrupolar topological insulators (QTI) have attracted significant interest as a prime example of HOTI, where the bulk invariant (quadrupole moment) leads to the accumulation of quantized charge at the zero-dimensional corners of a two-dimensional supercell \cite{PhysRevB.96.245115, doi:10.1126/science.aah6442, PhysRevB.101.195309}. 
These zero dimensional corner modes act as a signature of the unconventional bulk-boundary correspondence specific to the higher order phase.
It is to be noted that the quadrupole moment is a direct extension to higher order, of the Berry phase machinery that classifies band topology.
QTIs have been explored extensively under several backdrops, namely periodic drive, non-Hermiticity and random disorder. 
Floquet QTI, which exhibits the presence of corner modes characterized by a quantized floquet quadrupole moment has been recently studied \cite{hu2020dynamical, nag2019out}.
Furthermore, trivial lattices, when subjected to intricate external losses can give rise to quadrupole topology characterized by biorthogonal nested Wilson loop spectra \cite{tian2023analogous}. 
Additionally, in a pioneering work by Li \textit{et al.} \cite{PhysRevLett.125.166801}, it was shown that quadrupole topology can be successfully defined in the presence of disorder provided the system satisfies chiral symmetry.
This opened up a plethora of research on the interplay of disorder and topology in tight binding systems including the work done by Yang \textit{et al.}, where random onsite disorder revives the higher order topological phase in a quadrupolar insulator beyond its topological regime \cite{yang2021higher, yang2024higher}.
Motivated by this interplay caused by the conjunction between random disorder and topology we study the Benalcazar-Bernevig-Hughes model, which is a prototypical example of a quadrupolar insulator hosting zero-dimensional corner modes, under the action of a quasiperiodic (QP) disorder introduced in the hopping potential. 
Incommensurate or quasiperiodic potential constitutes a bridge between the completely periodic and completely random regime, thus manifesting a manifold of novel phases not exhibited by either limit \cite{PhysRevB.106.184209, PhysRevB.101.014205}.
While clean systems possessing translational invariance host periodic Bloch states, the presence of completely random disorder leads to absolute localization of states for arbitrarily small disorder strength in 1D and 2D.
However in 3D, the existence of a mobility edge is possible that represents a critical point of separation in energy between the localized and the extended states \cite{mott1987mobility}.
QP systems on the other hand can achieve such a transition even in one-dimension.
This was beautifully exhibited in the Aubrey-Andre (AA) model which represents one of the most popular examples of QP lattice \cite{dominguez2019aubry, biddle2009localization}.
It introduces an incommensurate onsite potential in a tight binding 1D chain with nearest neighbor hopping amplitudes.
The 1D AA model exhibits a complete localization-delocalization transition at a specific value of the disorder strength which was unforeseen in 1D system with random disorder.
It should be mentioned here that while higher order topology in a 2D Aubrey-Andre-Harper model has been studied \cite{PhysRevB.101.241104}, the complete effect of the quasiperiodic disorder on the topology of the system, including its capacity to restore a topological phase in the otherwise trivial regime, have not been explored in details.
Our primary motivation in this work has therefore been to address the effect of quasiperiodicity on the Benalcazar-Bernevig-Hughes (BBH) model, which is a well known quadrupolar insulator, and study the different topological phases that emerge due to this interplay.

The paper is structured as follows.
\begin{figure}
\centering
\includegraphics[width=0.85\columnwidth]{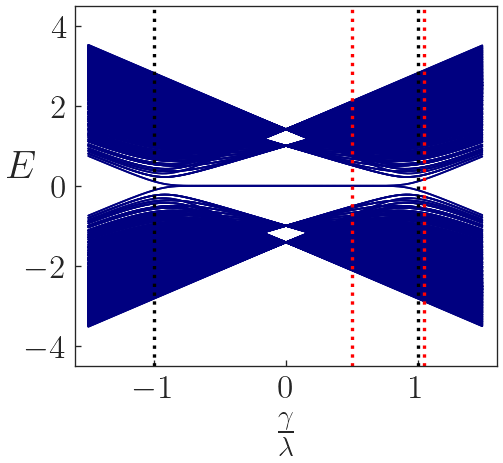}
\caption{\label{22222}The bandstructure of the original BBH model is plotted as a function of $\frac{\gamma}{\lambda}$. Robust mid band states, affixed at zero energy, are observed for $|\frac{\gamma}{\lambda}|<1$. The size of the lattice has been considered to be $40\times 40$. The black lines denote the quadrupolar topological regime, whereas the red lines mark the points $\frac{\gamma}{\lambda}=1.05$ and $\frac{\gamma}{\lambda}=0.5$, for which the effect of disorder on the topology of the system is studied.}
\end{figure}
In section II, we introduce the Hamiltonian of the original BBH model followed by a discussion on its topology and boundary modes.
In section III, we study the effect of the quasiperiodic disorder on the topology of the model, starting from a regime where the system is trivial in the clean limit.
In section IV, a similar analysis is done on the system, this time residing within the regime that is topological in the clean limit.
We also explore several indicators of the localization phenomena to decipher the confinement of the topological states.
Finally, we conclude in Section V.
\section{The Benalcazar-Bernevig-Hughes model}
The existence of a QTI requires at least two occupied bands in addition to the presence of crystalline symmetries that quantize the quadrupole moment \cite{doi:10.1126/science.aah6442}.
The BBH model, which is a four band insulator, abiding by the aforementioned criteria, is one of the most well-studied examples of a QTI.
The Hamiltonian for the BBH model is given as:
\begin{align}
\begin{split}
H(k_x, k_y)&=[\gamma + \lambda\cos(k_x)]\tau_1\sigma_0 -\lambda\sin(k_x)\tau_2\sigma_3 \\&- [\gamma + \lambda\cos(k_y)]\tau_2\sigma_2-\lambda\sin(k_y)\tau_2\sigma_1 + \delta\tau_3\sigma_0
\end{split}
\end{align}
\begin{figure} 
\centering
    \includegraphics[width=0.85\columnwidth]{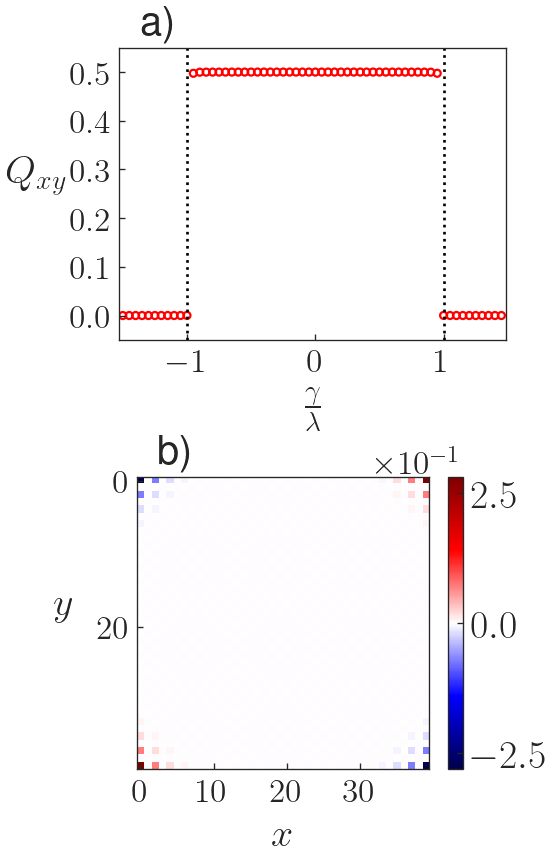}  
\caption{\label{3}(a) The quadrupole moment $Q_{xy}$ plotted as a function of $\frac{\gamma}{\lambda}$, shows a non-trivial value of $0.5$, for $|\frac{\gamma}{\lambda}|<1$. This confirms the bulk-boundary correspondence in the second order quadrupole topological insulator phase. (b) The excess charge distribution in the system for $|\frac{\gamma}{\lambda}|<1$, shows a unique pattern with an excess and deficit of charge accumulation at the alternate corners. More specifically, here $\gamma=0.5$ and $\lambda$ has been taken to be $1$ throughout.}
\end{figure}
The unit cell comprises of four sublattices as shown in Fig. \ref{1}. 
A magnetic field penetrates the system uniformly bestowing a total of $\pi$ flux per plaquette.
Here, $\gamma$ and $\lambda$ correspond to the intracell and intercell hoppping respectively.
$\delta$ corresponds to the onsite potential and is kept infinitesimally close to zero. 
Furthermore, the Pauli matrices $\tau$ and $\sigma$ correspond to the orbital degrees of freedom.
The bandstructure of the BBH model shows the existence of zero energy modes (Fig. \ref{22222}) characterized by a topological quadrupole moment for $|\frac{\gamma}{\lambda}|<1$ (Fig. \ref{3}(a)).
The quadrupole moment is protected by the dual mirror symmetry $M_{x}$ and $M_{y}$ such that,
\begin{equation}
M_{\alpha}H(k_x, k_y)M_{\alpha}^{\dagger}=H(M_{\alpha}\mathbf{k}), \text{\hspace{2mm}}\alpha\in x, y
\end{equation}
where, $M_x\mathbf k=(-k_x, k_y)$ and $M_y\mathbf k=(k_x, -k_y)$. 
At $\delta=0$ and $|\frac{\gamma}{\lambda}|<1$, four degenerate zero-energy in-gap corner states emerge.
On increasing the value of $\delta$ slightly, the degeneracy of the zero-energy states is lifted and the system exhibits a unique charge density pattern at the corners of a square supercell which is reminiscent of the bulk quadrupole topology (Fig. \ref{3}(b)).
The accumulated charge is quantized to a value of $\pm \frac{e}{2}$.
In the regime $|\frac{\gamma}{\lambda}|>1$, however, the zero energy corner states vanish and the second order topology in the system is lost. 
It should be mentioned here that for the emergence of a quantized quadrupole moment, it is necessary for the reflection operators, that is $M_x$ and $M_y$, to anticommute.
To calculate the quadrupole moment $Q_{xy}$, we now resort to the formalism prescribed in \cite{PhysRevB.100.245134} which provides an alternative approach as compared to the nested Wilson loop method.
In this method, the real space charge distribution is employed to evaluate $Q_{xy}$ as follows \cite{PhysRevLett.125.166801}:
\begin{equation}
    Q_{xy}=\frac{1}{2\pi}\mathrm{Im}\text{ }\mathrm{log}[\text{det}(U^\dagger\hat q U)\sqrt{\text{det}(\hat q^\dagger)}]
\end{equation}
Here, $\hat q=\mathrm{exp}[2\pi i\hat Q_{xy}]$, where $\hat Q_{xy}=\frac{\hat x\hat y}{L_xL_y}$. $\hat x$($\hat y$) and $L_x$($L_y$) correspond to the position operator and the length of the system respectively, in the $x$($y$) direction.
Additionally, the matrix $U$ is constructed by columnwise packing all the occupied eigenstates of the BBH Hamiltonian under periodic boundary conditions (PBC).
It is observed that the quadrupole moment exhibits a quantized value of $\frac{1}{2}$, for the regime $|\frac{\gamma}{\lambda}|<1$, while becoming trivial for $|\frac{\gamma}{\lambda}|>1$.
It should be mentioned here that a direct manifestation of $Q=\frac{1}{2}$, at the boundary is the accumulation of excess corner charges of magnitude $\pm\frac{e}{2}$.
This establishes the refined BBC in the second order topological phase.
\section{Disordered BBH Model}
\subsection{Clean Limit: Trivial Phase}
\subsubsection{Topological properties of the model}
\begin{figure}
\centering
\includegraphics[width=0.85\columnwidth]{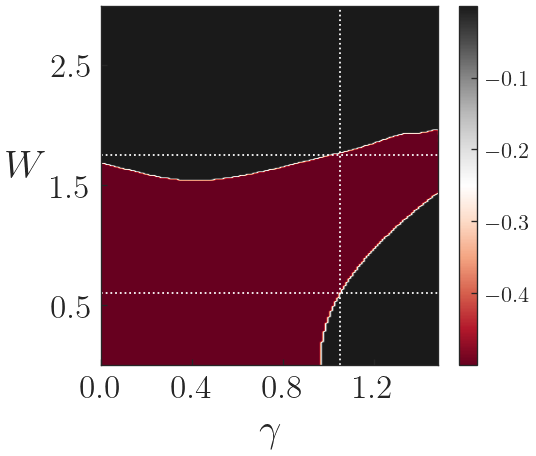}
\caption{\label{4}The variation of the quadrupole moment $Q_{xy}$ as a function of the disorder strength $W$ and the intracellular hopping parameter $\gamma$, is plotted. The vertical dotted line corresponds to $\gamma=1.05$, which represents the trivial limit in the clean case. The horizontal lines mark the regime where the system enters into a disorder driven second order topological phase.}
\end{figure}
Disorder possesses the capacity to generate novel topological phases when introduced in suitable systems.
To explore this interplay, we introduce a quasiperiodic disorder in the intracellular hopping amplitude of the BBH model such that:
\begin{align}
\begin{split}
&\gamma_{x_n} = \gamma + W\cos(2\pi\beta x_n + \phi)\\&\gamma_{y_n} = \gamma + W\cos(2\pi\beta y_n + \phi)
\end{split}
\end{align}
Here $\gamma_{x_n}$($x_n$) and $\gamma_{y_n}$($y_n$) correspond to the hopping amplitude(position coordinate) along the $x$ and $y$ directions respectively, for the $n^{\text{th}}$ lattice site.
The intercellular hopping amplitude $\lambda$ is however kept fixed at $1$.
$W$ corresponds to the strength of the disorder.
We have kept $\beta$ equal to $\frac{1}{\sqrt 2}$ which is responsible for the introduction of incommensurability in the system.
\begin{figure}
\centering
\includegraphics[width=0.85\columnwidth]{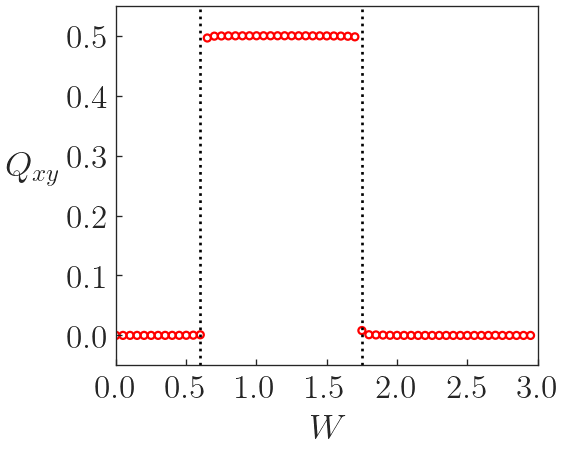}
\caption{\label{5}The quadrupole moment $Q_{xy}$ indicates a non-trivial topology in the regime $W\in(0.6, 1.75)$ thus proving the restoration of BBC by the quasiperiodic disorder. Here we have taken $\frac{\gamma}{\lambda}=1.05$, which corresponds to the trivial limit in the clean case.}
\end{figure}
\begin{figure}
\centering
\includegraphics[width=0.85\columnwidth]{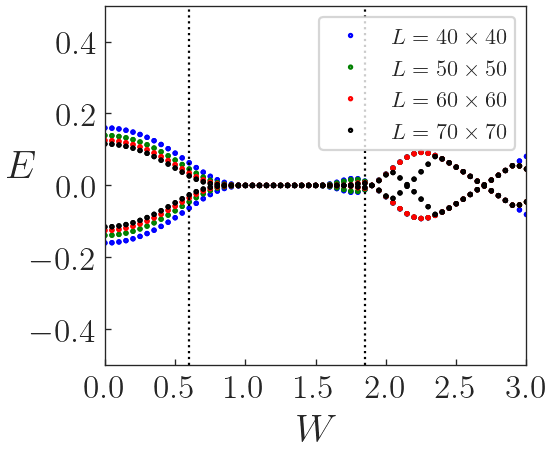}
\caption{\label{6}The energy eigenvalues of the mid-band states are plotted as a function of $W$. The dotted lines correspond to $W=0.6$ and $1.75$, which demarcate the disorder driven topological phase from the trivial one. We observe that the mid-band states are affixed at zero energy in the topological regime. Here, $\frac{\gamma}{\lambda}=1.05$.}
\end{figure}
\begin{figure*}
    \centering
    \includegraphics[width=2.2\columnwidth]{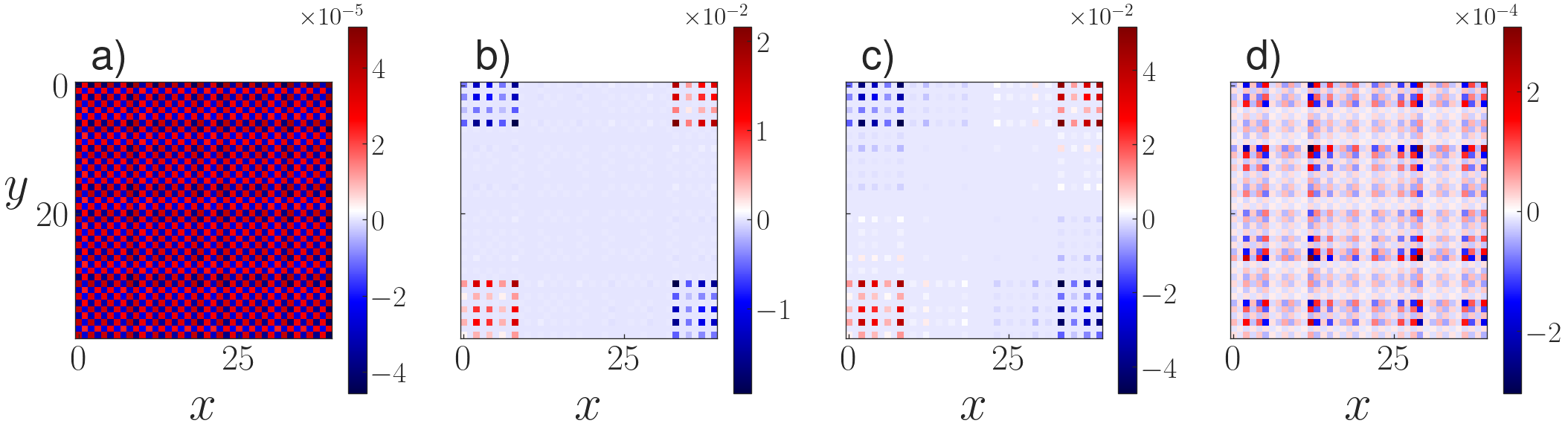} 
    \caption{\label{7}The excess charge distribution has been plotted for (a) $W=0.3$, (b) $W=1.0$, (c) $W=1.2$ and (d)$W=2.1$ respectively. It is clearly seen that the distribution shows no quadrupolar pattern in the trivial regime, that is $W<0.6$ and $W>1.75$, as shown in (a) and (d). However, a distribution pattern similar to Fig. \ref{3}(b) is seen in the topological regime ((b) and (c)).}    
\end{figure*}
$\phi$ corresponds to the phase of the disorder which is kept $0$ throughout.
It may be noted that different variants of the QP potential exist in literature, however the one considered here is the most common.
We plot the quadrupole moment $Q_{xy}$ as a function of the disorder amplitude $W$ and the intracellular hopping amplitude $\gamma$ (Fig. \ref{4}).
The invariant has been calculated on a $40\times 40$ lattice and a comparison between different system sizes has been done later.
It should be mentioned here that the quasiperiodic disorder has been introduced in the model such that the original chiral symmetry of the system, which is given by $\tau_3\otimes\sigma_0$, is kept intact.
This is crucial for the quantization of the quadrupole moment, as mentioned in \cite {PhysRevLett.125.166801}.
Furthermore, the nested Wilson loop approach fails here, owing to the lack of translational symmetry in the crystal due to the inclusion of disorder potential.
We observe in Fig. \ref{4} that the system undergoes a transition into the second order topological phase characterized by the quadrupole moment in the regime $\frac{\gamma}{\lambda}>1$ (which is completely trivial in the clean case), for a finite non-zero value of the disorder strength.
This implies that the quasiperiodic disorder restores the second order topology in the BBH model for a finite range of the disorder strength, prior to completely trivializing the system.
We now fix the value of the ratio $\frac{\gamma}{\lambda}$ at $1.05$, which corresponds to the trivial limit ($Q_{xy}=0$) in the clean case.
Fig. \ref{5} shows the plot of the quadrupole moment solely as a function of the disorder strength.
The plot shows a non-trivial value of $Q$ for $W$ ranging roughly from $0.6$ to $1.75$.
The bandstructure of the disordered BBH model under open boundary conditions (OBC) also shows the formation of zero energy states in this region (Fig. \ref{6}).
The agreement between the quadrupole moment and emergence of zero energy states thus confirms the restoration of the refined BBC in the disorder induced second order topological phase.
\begin{figure}
    \centering
    \includegraphics[width=0.8\columnwidth]{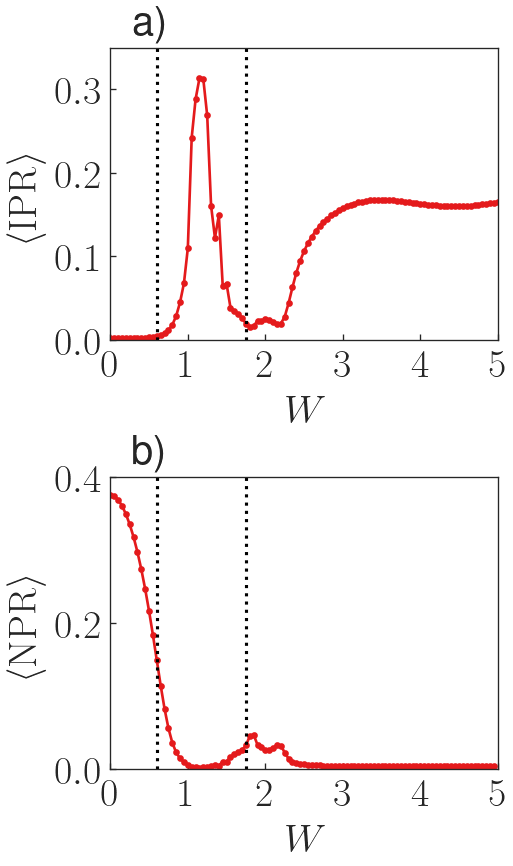}    
    \caption{Averages of IPR and NPR namely, \label{8}(a) $\langle \mathrm{IPR} \rangle$ and (b) $\langle \mathrm{NPR} \rangle$ are plotted as a function of the disorder strength. It is clear that below $W=0.6$, the mid gap states are extended, while being localized above that. However, around $W=1.75$, both $\langle \mathrm{IPR} \rangle$ and $\langle \mathrm{NPR} \rangle$ exhibit finite non-zero values, indicating at a possible multifractal state during phase transition. The two dotted lines demarcate the regime where the quadrupole moment is non-trivial.}
\end{figure}
Furthermore, the excess charge density is plotted for different values of the disorder strength corresponding to the trivial and topological phases (Fig. \ref{7}). 
It is observed that the accumulation of charge densities at the corners of a square supercell is negligible when $W>1.75$ and $<0.6$, indicative of a trivial topology.
However, for $0.6<W<1.75$, the distribution of excess charge at the corners shows a pattern which is suggestive of the inherent bulk quadrupole topology.
\subsubsection{Localization study}
In this section, we embark upon the localization properties of the disordered BBH model.
For this purpose, two diagnostic quantities, namely the inverse participation ratio ($\mathrm{IPR}$) and the normalized participation ratio ($\mathrm{NPR}$) are investigated upon.
The $\mathrm{IPR}$ can be mathematically represented as \cite{evers2000fluctuations}:
\begin{equation}
\mathrm{IPR}^m = \sum_{i=1, \alpha}^{{N}}|u_{i, \alpha}^m|^4,\text{\hspace{2mm}}\alpha\in{\text{A, B, C, D}}
\end{equation}
where $m$ denotes the band index and $N$ denotes the total number of lattice sites in the square supercell.
$|u^m\rangle$ corresponds to the $m^{\text{th}}$ eigenstate of the disordered BBH Hamiltonian under OBC.
$\mathrm{IPR}$ represents an important signature of localization in condensed matter systems which tends to $0$ for extended states while being finite for localized states (approaches $1$ in the thermodynamic limit).
The $\mathrm{NPR}$, on the other hand, can be represented as \cite{PhysRevB.107.014202}:
\begin{equation}
\mathrm{NPR} = \frac{1}{N}\frac{1}{\sum_{i=1, \alpha}^{{N}}|u_{i, \alpha}^m|^4},\text{\hspace{2mm}}\alpha\in{\text{A, B, C, D}}
\end{equation}
where the symbols represent the same quantities as mentioned earlier.
The $\mathrm{NPR}$ assumes a finite non-zero value in the extended case while tending to zero in the thermodynamic limit for the localized case.
More precisely, the behavior of $\mathrm{IPR}$ in localized and extended systems can be given as \cite{PhysRevB.101.064203}:
\begin{align}
\begin{split}
\mathrm{IPR}&\sim\frac{1}{L^D}\text{\hspace{3.5mm} : extended systems}\\&\sim
\mathcal{O}(1)\text{\hspace{2mm} : localized systems}
\end{split}
\end{align}
\begin{figure}
\centering
\includegraphics[width=0.85\columnwidth]{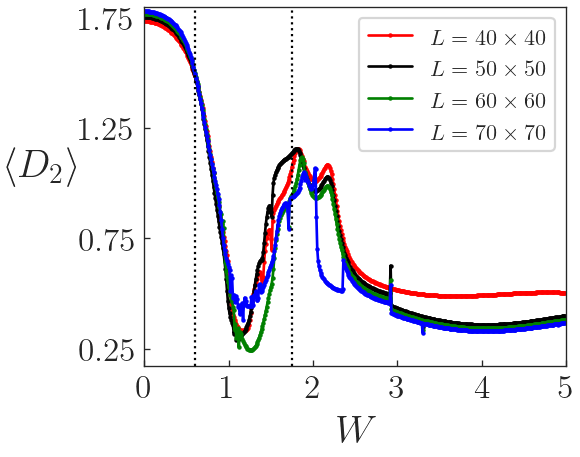}
\label{9}
\caption{\label{9} The average fractal dimension $\langle D_2\rangle$ exhibits a value close to $D=2$, for $W<0.6$, representing a trivial delocalized phase. In the topological regime, however, it approaches zero. During the phase transition (around $W=1.75$), $\langle D_2\rangle$ assumes a value between zero and $2$, thus providing concrete evidence of a multifractal phase. The calculations have been repeated for different system sizes to eliminate any possible finite size effect, as shown in the figure.}
\end{figure}
However, $\mathrm{NPR}$ shows a contrasting feature in the two regimes:
\begin{align}
\begin{split}
\mathrm{NPR}&\sim\mathcal{O}(1)\text{\hspace{2mm}: extended systems}\\&\sim
\frac{1}{L^D}\text{\hspace{3.5mm}: localized systems}
\end{split}
\end{align}
We plot the average value of $\mathrm{IPR}$ and $\mathrm{NPR}$ (denoted by $\langle \mathrm{IPR}\rangle$ and $\langle \mathrm{NPR}\rangle$) for the four degenerate zero energy states (Fig. \ref{8}).
It is observed that the values of $\langle \mathrm{IPR}\rangle$ remain close to zero till roughly $W=0.6$, beyond which it assumes a finite non-zero value. 
Around $W=1.75$, the value of $\langle\mathrm{IPR}\rangle$ decreases, before finally increasing again.
This indicates that the four mid-band states are delocalized when $W<0.6$, which is similar to what happens in the clean case for $|\frac{\gamma}{\lambda}|>1$.
Beyond $W=0.6$, the disorder driven quadrupole phase affixes these mid-band states at zero energy and localizes them.
For higher values of the disorder strength (beyond $W=1.75$), the states are no longer at zero energy and are not topological.
The decrease in the value of $\mathrm{IPR}$ around $W=1.75$, indicates a delocalization caused by the phase transition that destroys the quadrupole topology in the system.
Much beyond $W=1.75$, localization sets in due to the larger values of the disorder strength.
The system is then in a trivial Anderson localized phase.
The variation of $\langle \mathrm{NPR}\rangle$ as a function of $W$ (Fig. \ref{8}(b)) provides similar information.
\begin{figure}
\centering
\includegraphics[width=0.8\columnwidth]{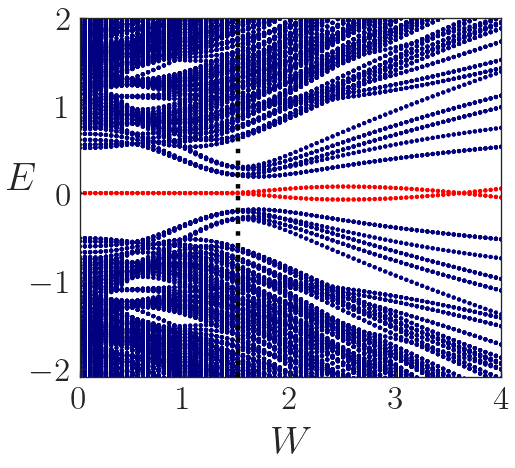}
\caption{\label{10}The bandstructure of the BBH model as a function of the disorder strength $W$ for $\frac{\gamma}{\lambda}=0.5$. It is observed that the in-gap states, that are depicted in red, remain confined at zero energy as long as $W\leq 1.5$.}

\end{figure}
\begin{figure}
\centering
\includegraphics[width=0.8\columnwidth]{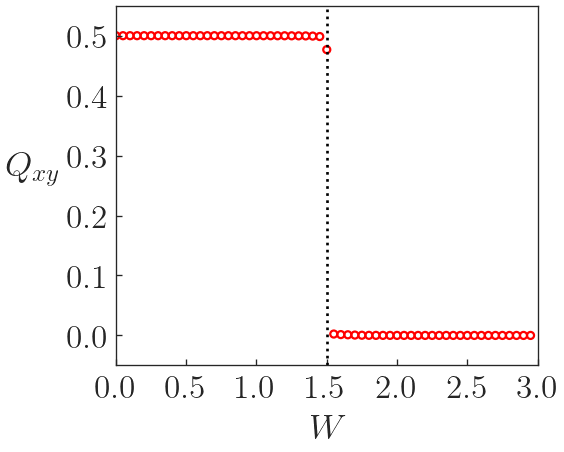}

\caption{\label{11}The quadrupole moment $Q_{xy}$ shows a behavior that accurately reciprocates the behavior of the bandstructure in Fig. \ref{10}. $Q_{xy}$ remains affixed at $0.5$, indicating at a topological phase, for the regime $W\leq 1.5$. Beyond this point, the system gets trivialized.}
\end{figure}
\begin{figure}
\centering
\includegraphics[width=0.75\columnwidth]{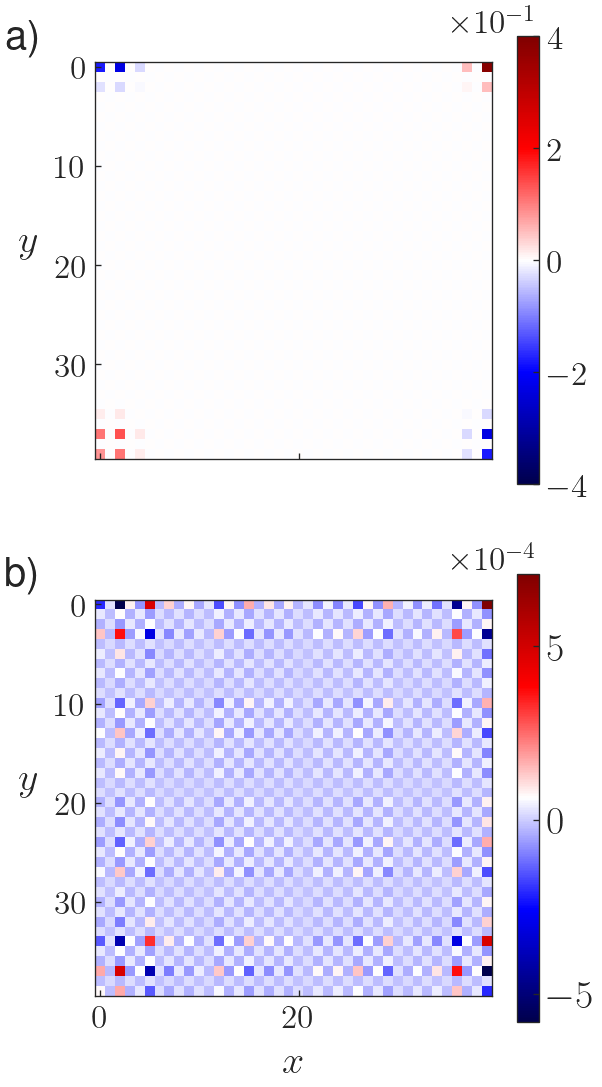}
\caption{\label{12}The distribution of excess charge on a square supercell of size $40\times40$ for (a) $W=1.0$ and (b) $W=1.8$. Evidently for $W\leq 1.5$, the system persists in the second order quadrupolar phase.}
\end{figure}
\begin{figure}
\centering
\includegraphics[width=0.85\columnwidth]{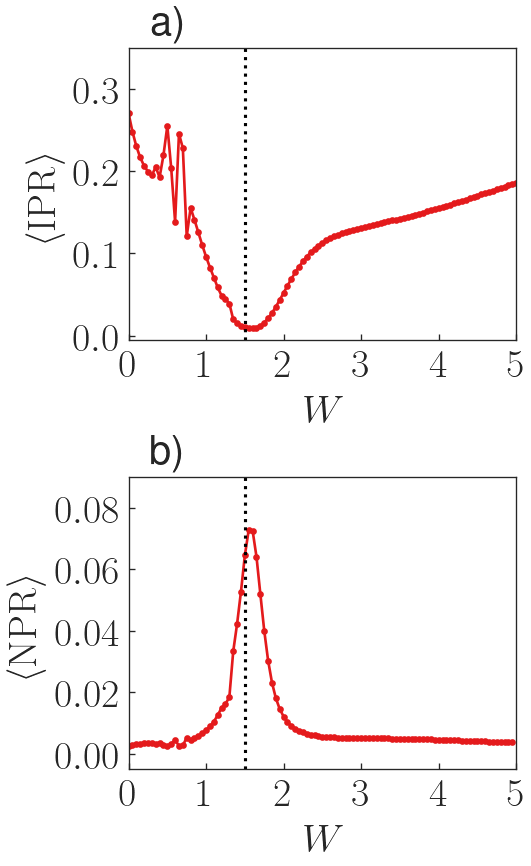}
\caption{\label{13}The variation of (a) $\langle \mathrm{IPR}\rangle$ and (b) $\langle \mathrm{NPR}\rangle$ shows that the system is localized both in the topological ($W\leq 1.5$) and in the trivial ($W>1.5$) phase. However, the dip(spike) in the value of $\langle \mathrm{IPR}\rangle$($\langle \mathrm{NPR}\rangle$) around $W=1.5$ hints at a topological phase transition.}
\end{figure}

Another widely investigated concept in the study of localization is the fractal dimension. 
Inspite of $\mathrm{IPR}$ and $\mathrm{NPR}$ being excellent signatures of localization/delocalization in a system, they are not individually efficient indicators of multifractality.
Therefore, to investigate upon the existence of possible multifractality in the system, we calculate the fractal dimension $D_2$ which assumes a value close to $D$ in the thermodynamic limit for the extended states, while being zero for the localized states. 
However, in the multifractal phase, $D_2\in(0,D)$.
Here $D$ refers to the physical dimensionality of the system which is $2$ in our case.
The mathematical representation of $D_2$ can be given as \cite{PhysRevB.107.014202}:
\begin{equation}
D_2=-\lim_{L\to\infty}\frac{\log(\mathrm{IPR})}{\log(L)}
\label{E9}
\end{equation}
Here $L=L_x=L_y$ refers to the length of the system along the $x$ or $y$ direction.
We again plot the average value of $D_2$ for the four mid-gap states, as a function of the disorder strength in Fig. \ref{9}.
$\langle D_2\rangle$ shows a value close to $2$ and $0$ for the regime $W<0.6$ and $W>0.6$ respectively, which is consistent with the extended and localized nature of the mid-gap states.
However, around $W=1.75$, where the topological phase transition occurs, the value of $\langle D_2\rangle$ lies distinctly in between the two extremes, that is $0<D_2<D$.
Thus it is evident, that close to the transition from a topological to a trivial phase, the zero energy energy states acquire a fractal nature before getting completely localized again.
The calculations have been repeated for various system sizes and they provide robust evidence to the inferences made above.
\subsection{Clean Limit: Topological Phase}
\subsubsection{Topological properties}
We now focus on the effect of disorder on the topological properties of the BBH model, starting within a regime which is non-trivial in the clean limit. 
We fix the value of $\frac{\gamma}{\lambda}$ at $0.5$ which originally corresponds to a quadrupolar topological insulator (Fig. \ref{22222}(a)).
The bulk bandstructure is similarly plotted as a function of the disorder strength $W$ (Fig. \ref{10}).
We observe that the in-gap states remain confined to zero energy till $W=1.5$, beyond which they shift, indicating a possible destruction of the topological feature due to the imposed disorder.
This is further confirmed by the quadrupole moment, which shows a transition to triviality at $W=1.5$ (Fig. \ref{11}).
Furthermore, the distribution of the excess charge at $W=1.0$ shows charge accumulation of $\pm \frac{e}{2}$ at alternate corners similar to the previous case (Fig. \ref{12}(a)).
Beyond $W=1.5$, (as shown in Fig. \ref{12}(b) for $W=2.1$) there is negligible charge accumulation in the system, which is expected from a system with trivial quadrupole moment (Fig. \ref{12}).
All the calculations have been done on a $40\times40$ square supercell.
\begin{figure}
\centering
\includegraphics[width=0.85\columnwidth]{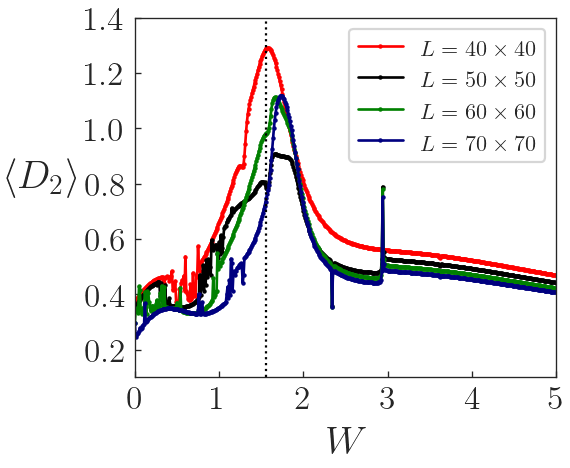}
\caption{\label{14}The fractal dimension $D_2$ shows that the in-gap states exhibit a multifractal nature near the transition point $W=1.5$, while being localized both in the trivial and the topological phase.}
\end{figure}
\subsubsection{Localization properties}
To study the localization of the in-gap states we again resort to a study of $\mathrm{IPR}$ and $\mathrm{NPR}$ as a function of $W$ (Fig. \ref{13}).
Interestingly, the system persists in a localized regime deep within the topological as well as the trivial phase.
However, near the transition point, the states tend towards delocalization.
This is exhibited by a sudden dip(spike) in the value of $\langle\mathrm{IPR}\rangle$($\langle\mathrm{NPR}\rangle$) around the transition point at $W=1.5$.
Furthermore, finite and non-zero values of both $\langle\mathrm{IPR}\rangle$ and $\langle\mathrm{NPR}\rangle$ around the transition point indicate that the in-gap states might show possible multifractality.
To confirm this, we resort to the evaluation of the fractal dimension $D_2$ (Eq. \ref{E9}).
We observe that $D_2$ exhibits a value close to $0.4$, indicating at a localized phase on both sides away from the transition point (Fig. \ref{14}). 
However, around the transition point $D_2\in(0, 2)$,
more specifically $D_2\sim 1.0$ near $W=1.5$, thus indicating at a multifractal phase.
\section{Conclusion}
We observe a topological phase transition induced purely by a quasiperiodic disorder inculcated in the hopping amplitude of the Benalcazar-Bernevig-Hughes model.
The quadrupole moment shows a restored topological phase as a function of the disorder strength, in the regime where the clean system shows completely trivial behavior.
The bandstructure of the BBH model under OBC also shows the emergence of zero energy in-gap states in exact correspondence with the quadrupole moment, thus providing clear evidence of a second order topological phase.
Furthermore, the distribution of charge at the corners of the square supercell, also shows a pattern that is suggestive of the bulk quadrupole topology.
A localization study done on the mid-band states shows three different phases namely: trivial delocalized, topological localized and Anderson localized phase.
The fractal dimension $D_2$ shows that near the transition from topological localized to trivial localized phase, the system becomes multifractal which is characterized by a value of $D_2$ in between zero and $D$.
A similar phenomena occurs, when we study the system as a function of disorder starting from a clean limit that is topological. 
Here, the disorder drives the in-gap states from being topologically localized to Anderson localized corresponding to higher values of the disorder strength $W$.
The point of transition is again host to a multifractal phase characterized by finite non-zero values of both $\langle \mathrm{IPR}\rangle$ and $\langle \mathrm{NPR}\rangle$.
The fractal dimension $D_2$ provides further concrete evidence for this multifractality.
\bibliographystyle{apsrev4-2}
\bibliography{ref}
\end{document}